\documentclass[tightenlines,twocolumn,superscriptaddress,preprintnumbers,amssymb,nofootinbib]{revtex4-2}
\usepackage{amsmath, amsthm, amssymb, mathrsfs, bm, mathtools} 
\usepackage{graphics}
\usepackage{epsfig}
\usepackage{graphicx}
\usepackage{wrapfig}
\usepackage{color}
\usepackage{tikz}
\usepackage[hidelinks,colorlinks=true,linkcolor=blue,citecolor=blue]{hyperref}

\RequirePackage{subfigure}

\newcommand{\beq}{\begin{equation}}
\newcommand{\eeq}{\end{equation}}
\newcommand{\bqa}{\begin{eqnarray}}
\newcommand{\eqa}{\end{eqnarray}}

\newcommand{\ms}{\overline{\text{\tiny MS}}}

\newcommand{\x}{x}
\newcommand{\y}{y}

\begin{document}

\title{ Consistent chiral limit in on-shell renormalized  quark-meson model with ChPT  scaling}
\author{Vivek Kumar Tiwari}
\email{vivekkrt@gmail.com}
\affiliation{Department of Physics, University of Allahabad, Prayagraj, India-211002}
\date{\today}

\begin{abstract}

Chiral perturbation theory (ChPT) predicted scaling of the pion,~kaon decay  constants $f_{\pi}, f_{K} $ and $M_{\eta}^2 = m_{\eta}^2 + m_{\eta^{\prime}}^2 $ has been used in the renormalized $2+1$ flavor quark-meson (RQM) model,~to find a consistent path to chiral limit as  the $\pi  \ \text{and} \ K $ meson masses approach zero.~The left side of the Columbia plot is generated free from any ambiguity or heuiristic adjustment in  the model parameter fixing  away from the physical point.
The on-shell renormalization of parameters  and consistent treatment of the  divergent quark one-loop vacuum fluctuations in the RQM model,~make the axial $U_{A}(1)$ anomaly  significantly stronger and the light (strange) explicit chiral symmetry breaking strength becomes weaker by a small (relatively large) amount.~The first order chiral transition region in the $m_{\pi}-m_{K}$ and $\mu-m_{K}$ planes of Columbia plot,~increases due to the above novel features of the RQM model.~Comparing the softening effect of quark one-loop vacuum fluctuation on the chiral transition,~it looks  overestimated in a recent functional renormalization group study.
  
\end{abstract}
\keywords{ Dense QCD,
chiral transition,}
\maketitle

The chiral phase transition is the focus of intensive research since the universality \cite{rob} predictions that it is first order in the chiral limit of $u,d,s$ quark masses $m_{u}=m_{d}=0=m_{s}$ and second order of the $O(4)$ universality class if $m_{u,d}=0$, $m_{s}=\infty$ with the strong $U_A(1)$ anomaly \cite{tHooft:76prl} ($\eta^{\prime}$ mass $m_{\eta^{\prime}}(T_{c})>>T_{c}$) at the chiral symmetry restoring  critical temperature $T_{c}$.~The second order tansition turns first order at a triple point for some finite $m_{s}=m_{s}^{TCP}$($m_{K}=m_{K}^{TCP}$) in the light chiral limit $m_{u,d}=0$ ($m_{\pi}=0$).~The mass dependence of the order of phase transition,~structures of the critical lines and surfaces at zero, real/imaginary chemical potentials $\mu$,~is summarized in the Columbia plot \cite{columb} which is well understood in the heavy quark mass limit  both from the continuum and lattice QCD (LQCD) studies \cite{from, saito, reinosa, fischrA, mael}.~The LQCD has settled that the chiral symmetry restoration at physical point for $\mu=0$,~is a crossover transition with pseudocritical temperature $T_{\chi}\equiv 155\pm2$ MeV \cite{Wupertal2010, WB2010II, HotQCD2012, WB2014, HotQCD2014}.

The precise mapping of the critcical line separating the region of  first order transition from that of the crossover for small masses in the $m_{u,d}-m_{s}$ ($m_{\pi}-m_{K}$) plane,~is quite challenging in the LQCD \cite{karsch1, karsch2, karsch3, forcrd, varnho, jin, Bazav, jin2} as the results show large variations due to  the strong cutoff and discretization effects.~The crossover at $\mu=0$ turns first order for the critical quark (pion) mass $m_{q} \ (m_{\pi})<m_{q}^{c} (m_{\pi}^{c})$.~For $N_{f}=3$ degenerate quarks,~the LQCD studies conducted between 2001 to 2017,~confirm the first order region close to the chiral limit and find the  $m_{\pi}^{c}$ in the range 290-50 MeV \cite{forcrnd2,Resch}.~A recent improved LQCD $N_{f}=3$ study \cite{dini21} gets no direct evidence of first order transition for 80 MeV $\le m_{\pi} \le$ 140 MeV.~Using Mobius domain wall fermions,~the Ref.\cite {zhang24} finds small $m_{q}^{c}\le 4$ MeV while another LQCD chiral limit study suggests that the chiral transition could be of second order.~Although the improved LQCD studies,~suggest a small or no first order region in some cases,~the picture is not very clear as the resluts still show huge discrepancies because of the notoriously difficult problem of implementing chiral fermions on the lattice.~Apart from LQCD, a $N_{f}=3$ study in Dyson-Schwinger approach \cite{bernhardt23} and another in ef.\cite{kousvos22} finds  the second order transition.~Very recent functional renormalization group (FRG) study in local potential approximation (LPA) \cite{fejos22, fejoHastuda},~showed that in contrast to the $\epsilon$ expansion prediction\cite{rob},~the transition can be of second order for $N_{f}\ge2$ if the $U_A(1)$ symmetry gets restored at the $T_{\chi}$.~But they have added a note of caution for drawing final conclusions as the LPA completely neglects the wave function renormalization.~The thermal fate of the axial $U_A(1)$ anomaly is still unsettled \cite{lahiri21} as some studies  find that it is relevant at the critical point $T_{\chi}$ \cite{dick15, ding21, kaczmarek21, bazazov12, bha14, kaczmarek23} while others \cite{dini21, brandt16, tomiya16, aoki21, aoki22} claim that it vanishes.~Novel signals if anomaly strength becomes very weak at $T_{\chi}$,~has  been  conjectured in very recent  Ref.\cite{pisarski24}.

The effective model studies of QCD phase transition,~like,~linear sigma model (LSM) \cite{Ortman, Lenagh, Rischke:00, Roder, jakobi, Herpay:05, Herpay:06, Kovacs:2007,Jakovac:2010uy, marko, Fejos},~quark-meson (QM) \cite{scav, mocsy, bj, Schaefer:2006ds, SchaPQM2F, Bowman:2008kc,  SchaeferPNP, Schaefer:09,  Schaefer:09wspax, SchaPQM3F, Mao, TiPQM3F,  koch, zacchi1, zacchi2} or  Nambu-Jona-Lasinio\cite{costaA, costaB, fuku08} (NJL)  and the  non-perturbative FRG technique models 
\cite{berges, bergesRep, Gies, braunii, pawlanal, fuku11, Herbst, grahl, mitter, Weise1, FejosI, Renke2, brauniii, FejosII,FbRenk, Weise2, Tripol, Fejos3, Weise3, Fejos4, fejos5}~have a long history of complementing the LQCD studies with new insights.~Refining the QM model into a valuable tool of chiral limit studies,~one faces the challenge of 
finding the quark ( pion and kaon) mass dependence of the model parameters.~The often used method \cite{Ortman, Lenagh, Schaefer:09,  fuku08, berges, Herbst}  called the fixed-ultraviolet (UV) scheme \cite{Resch} relies on changing the light (strange) explicit chiral symmetry breaking strengths  $ h_{x}(h_{y}) $ while all other parameters are kept same as the ones at physical point.~In this scheme,~chiral limit can be explored only for unphysically large scalar $\sigma$ masses $m_{\sigma}\ge800$ MeV because the spontaneous chiral symmetry breaking (SCSB) gets lost in the chiral limit as the mass parameter $m^2$ turns positive for the realistic $m_{\sigma}=400-600$ MeV \cite{Resch,    Schaefer:09}.~In a recent FRG QM model study under LPA of the Columbia plot,~Resch et. al.\cite{Resch} proposed ChPT motivated fixed-$f_{\pi}$ scheme where they heuiristically adjust the initial effective action  to the larger scales ($\Lambda^{\prime} > \Lambda $) for every smaller mass in the path to chiral limit such that the $f_{\pi}$ always retains its physical value and hence the SCSB is not lost.~The changing scale ($\Lambda^{\prime} > \Lambda $) accounts for the change in parameters (same as at the physical point) when the strenghts $h_{x}(h_{y})$ decrease.

The present study aims to generate the left corner of the Columbia plot
and find a consistent path to chiral limit using the exact chiral effective  potential of the RQM model \cite{vkkr22, skrvkt24} whose parmeters are renormalized on-shell after the consistent treatment of the quark one-loop vauum fluctuation.~Using the $ \mathcal{O}(\frac{1}{f^2})$ \cite{gasser, herrNPB, herrPLB, borasoyI, borasoyII} accurate results of the ChPT,~first proposed in Ref.\cite{Herpay:05},~for the $ (m_{\pi},m_{K}) $ dependence of the tree level parameters,~the parameter fixing away from the physical point has been kept free from any ambiguity and heuiristic adjustment.~After renormalization,~the axial $U_{A}(1)$ anomaly  gets  significantly stronger in the RQM model while the strength  $h_{x}(h_{y})$ at the physical point gets reduced by a small (relatively large) amount because the pion(kaon)  curvature mass  $m_{\pi,c}(m_{K,c})$=135.95(467.99) MeV turns out to be 2.05(28.01) MeV smaller than its pole mass $m_{\pi}(m_{K})$=138(496) MeV.~In contrast,~the $c$ and $h_{x}(h_{y})$ do not change from their tree level values at the physical point in the FRG study under LPA where the wave function renormalization is completely neglected and the initial conditions (action of the low energy model at scale $\Lambda$) are not uniquely determined by solvig the RG evolution of QCD flow starting from a microscopic QCD action at a perturbatively large energy scale $k>>1$ GeV.~The tree level anomaly strength $c$ also increases towards the chiral limit 
due to the ChPT scaling of $f_{\pi}, f_{K} $ and $M_{\eta}$.~We will explore how the abovementioned new featurs of the RQM model,~impact the nature and extent of first or second order chiral transition in the Columbia plot with and without the $U_{A}(1)$ anomaly and compare the results with the e-MFA FRG study of \cite{Resch}.

{\bf {The RQM model}}- Lagrangian \cite{Rischke:00,Schaefer:09,TiPQM3F} is 
\bqa
\label{lag}
 {\cal L_{QM}}&=&\bar{\psi}[i\gamma^\mu D_\mu- g\; T_a\big( \sigma_a + i\gamma_5 \pi_a\big) ] \psi+\cal{L(M)}\;. \\  \nonumber
\label{lagM}
{\cal L(M)}&=&\text{Tr} (\partial_\mu {\cal{M}}^{\dagger}\partial^\mu {\cal{M}}-m^{2}({\cal{M}}^{\dagger}{\cal{M}}))\\ \nonumber
&&-\lambda_1\left[\text{Tr}({\cal{M}}^{\dagger}{\cal{M}})\right]^2-\lambda_2\text{Tr}({\cal{M}}^{\dagger}{\cal{M}})^2\\ 
&&+c[\text{det}{\cal{M}}+\text{det}{\cal{M}}^\dagger]+\text{Tr}\left[H({\cal{M}}+{\cal{M}}^\dagger)\right]\;.
\eqa
The Yukawa coupling $g$ couples the 3 flavor of quark fields $\psi$ (color $N_c$) to the nine scalar(pseudoscalar) meson fields $\sigma_a (\pi_a$) of $3\times3$ complex matrix ${\cal{M}}=\frac{\lambda_{a}}{2}(\sigma_{a}+i\pi_{a})$.~$\lambda_a$ ($a=0,1..8$) are Gell-Mann matrices,~$\lambda_0=\sqrt{\frac{2}{3}}{\mathbb I}_{3\times3}$
.~Nonzero condensates $\bar{\sigma_0}$ and $\bar{\sigma_8}$ break the $SU_L(3) \times SU_R(3)$ chiral symmetry spontaneously   while the external fields $H= T_{a} h_{a}$ with $h_0$, $h_8  \neq 0$ break it explicitly.~The change from the singlet octet $(0,8)$  to the nonstrange strange basis $(\x,\y)$ gives $\x ( h_{x})= \sqrt{\frac{2}{3}}\bar{\sigma}_0 (h_{0}) +\frac{1}{\sqrt{3}} \bar{\sigma}_8(h_{8})$ and $ \y (h_{y}) = \frac{1}{\sqrt{3}}\bar{\sigma}_0 (h_{0})-\sqrt{\frac{2}{3}}\bar{\sigma}_8 (h_{8})$.~With mesons as mean fileds and quarks-antiquarks with their thermal and quantum fluctuations,~the grand potential \cite{Schaefer:09, TiPQM3F} $\Omega_{\rm MF }(T,\mu)=U(\x,\y)+\Omega_{q\bar{q}} (T,\mu;\x,\y)\;$; the $ U(\x,\y) $ is vacuum effective potential and    $\Omega_{q\bar{q}}=\Omega_{q\bar{q}}^{vac}+\Omega_{q\bar{q}}^{T,\mu}$ is the quark/antiquark  contribution at  temperature $T$ and quark chemical potential $\mu_{f} (f=u,d,s)$.
\vspace {.0cm}
\bqa  
\label{eq:mesop}
\nonumber
\resizebox{0.9\hsize}{!}{$U(\x,\y)=\frac{m^{2}}{2}\left(\x^{2} +
  \y^{2}\right) -h_{x} \x -h_{y} \y
 - \frac{c}{2 \sqrt{2}} \x^2 \y$}\\  
\resizebox{0.9\hsize}{!}{$+ \frac{\lambda_{1}}{2} \x^{2} \y^{2}+
  \frac{1}{8}\left(2\lambda_{1}\lambda_{2}\right)\x^{4}+\frac{1}{8}\left(2 \lambda_{1} + 2\lambda_{2}\right) \y^{4} \; . $}\\
\label{vac1}
\resizebox{0.9\hsize}{!}{$\Omega_{q\bar{q}}^{vac} =- 2 N_c\sum_f  \int \frac{d^3 p}{(2\pi)^3} E_q \theta( \Lambda_c^2 - \vec{p}^{2})\; .$}\\
\label{vac2}
\resizebox{0.9\hsize}{!}{$\Omega_{q\bar{q}}^{T,\mu}=- 2 N_c \sum_{f=u,d,s} \int \frac{d^3 p}{(2\pi)^3} T \left[ \ln g_f^{+}+\ln g_f^{-}\right] \; . $} \\ 
\label{GrandQM}
\resizebox{0.9\hsize}{!}{$\Omega_{\rm QM }(T,\mu,x,y)=U(\x,\y)+\Omega_{q\bar{q}}^{T,\mu} \; .$} 
\eqa
The $ g^{\pm}_f = \left[1+e^{-E_{f}^{\pm}/T}\right] $; $E_{f}^{\pm} =E_f \mp \mu_{f}$ and $E_f=\sqrt{p^2 + m{_f}{^2}}$ is the quark/antiquark energy.~The light (strange) quark mass  $m_{u/d}=\frac{g\x}{2}$ ($m_{s}=\frac{g\y}{\sqrt{2}}$) and $\mu_{u}=\mu_{d}=\mu_{s}=\mu$.~The Eq.~(\ref{GrandQM}) is the grand  potential of QM model in the standard mean field approximation (s-MFA) 
where  quark one-loop vacuum term of Eq.~(\ref{vac1}) is dropped.

Several studies do the proper dimensional regularization of the divergences after including the vacuum fluctuations in the extended mean field approximation (e-MFA) \cite{ vac, lars, schafwag12, chatmoh1,  guptiw, vkkr12, vkkt13, Rai} but the effective potential $\Omega_{vac} (\x,\y)=U(\x,\y)+\Omega_{q\bar{q}}^{vac}$
in their treatment,~turns incosistent when they fix the model parameters 
using curvature masses of mesons whose self energy corrections are 
evaluated at the zero momentum \cite {laine, Adhiand1, BubaCar, fix1,  Adhiand2, Adhiand3, asmuAnd, RaiTiw, raiti2023} not in  the on-shell conditions.
~Here,~we will use the consistent e-MFA RQM model effective potential   calculated in our very recent works \cite{vkkr22, skrvkt24}  after relating the counter-terms in the $\overline{\text{MS}}$ scheme to those in the on-shell (OS) scheme \cite{Adhiand1, Adhiand2, Adhiand3, asmuAnd, RaiTiw, raiti2023} and finding  the renormalized parameters when the pole masses of the $m_{\pi}, m_{K}, m_{\eta},m_{\eta^{\prime}} \ \text{and} \ m_{\sigma}$,~the pion and kaon decay  constants $f_{\pi}$ and $f_{K}$ ,~are put into the relation of the running couplings and mass parameter.~In the $\overline{\text{MS}}$ scheme $ \Omega_{vac}=U(x_{\ms},y_{\ms})+\Omega^{q,vac}_{\ms}+\delta U(x_{\ms},y_{\ms})$ is rewritten in \cite{vkkr22} in terms of the scale $\Lambda$ independent parameters $\Delta_{x}=\frac{g_{\ms} \ x_{\ms}}{2}$ and $\Delta_{y}=\frac{g_{\ms} \ y_{\ms}}{\sqrt{2}}$ after 
$ 1 / \epsilon$ divergences gets cancelled.~The position of effective potential minimum is kept unchanged to fix the scale $\Lambda_0$.
~Although the $f_{\pi}$, $f_{K}$ and $g$ get renormalized due to the dressing of the meson propagator,~they do not change as $g_{\ms}=g_{ren}=g$, 
$x_{\ms}=x$, $y_{\ms}=y$ at $\Lambda_0$.~Further  $x_{\ms}=f_{\pi,ren}=f_\pi$ and $y_{\ms}=\frac{2f_{K,ren}-f_{\pi,ren}}{\sqrt{2}}$=  $\frac{2f_K-f_\pi}{\sqrt{2}}$ at the minimum.~The vacuum effective potential of \cite{vkkr22} can be  written back in terms of $x$ and $y$ as :
\bqa
\label{vacRQM}
\nonumber
&&\Omega_{vac}^{\rm RQM}(\x,\y)=\frac{(m^{2}+m^2_{\text{\tiny{FIN}}})}{2} \ \left(\x^{2} +
  \y^{2}\right)-(h_{x}+h_{x\text{\tiny{FIN}}}) \ \x 
  \\ \nonumber 
 &&-(h_{y}+h_{y\text{\tiny{FIN}}}) \y -\frac{(c+c_{\text{\tiny{FINTOT}}})}{2 \sqrt{2}} \x^2 \y + \frac{(\lambda_{1}+\lambda_{1\text{\tiny{FIN}}})}{2} \x^{2} \y^{2}
   \\ \nonumber
 &&+\frac{\left\{2(\lambda_{1} +\lambda_{1\text{\tiny{FIN}}})+
   ( \lambda_{2} +\lambda_{2\text{\tiny{FIN}}}  )\right\}\x^{4}}{8}+\left( \lambda_{1} +\lambda_{1\text{\tiny{FIN}}}+\lambda_{2}
 \right.  \\  \nonumber
 && \left. +\lambda_{2\text{\tiny{FIN}}}\right)\frac{ \y^{4}}{4} +\frac{N_c g^4 (\x^4+2\y^4)}{8(4\pi)^2} \left[\frac{3}{2}-\mathcal{C}(m^2_\pi)-m^2_\pi \mathcal{C}^{\prime}(m^2_\pi)\right]\ \\
 &&-\frac{N_c g^4 }{8(4\pi)^2} \left[\x^4 \ln\left(\frac{\x^2}{f_{\pi}^2}\right)+2\y^4 \ln\left(\frac{2 \  \y^2}{f_{\pi}^2}\right) \right]. \\ \nonumber  \\ 
\label{grandRQM}
&&\Omega_{\rm RQM }(T,\mu,x,y)=\Omega_{vac}^{\rm RQM}(\x,\y)+\Omega_{q\bar{q}}^{T,\mu}\;. 
\eqa 
The terms $\mathcal{C}(m^2_\pi) $ and $ \mathcal{C}^{\prime}(m^2_\pi) $ and the derivations for the renormalized parameters $m^{2}_{0}=(m^{2}+m^2_{\text{\tiny{FIN}}})$,~$h_{x0}=(h_{x}+h_{x\text{\tiny{FIN}}})$,~$h_{y0}=(h_{y}+h_{y\text{\tiny{FIN}}}) $,~$\lambda_{10}=(\lambda_{1}+\lambda_{1\text{\tiny{FIN}}})$,~$\lambda_{20}=(\lambda_{2} +\lambda_{2\text{\tiny{FIN}}})$ and $c_{0}=(c+c_{\text{\tiny{FINTOT}}})$ are given in detail in Refs.\cite{vkkr22}.~The experimental values of pseudoscalar meson masses $m_{\pi}$, $m_{K}$, $m_{\eta}$, $m_{\eta^{\prime}} $ ($m_{\eta}^2+m_{\eta^{\prime}}^2$), the scalar $\sigma$ mass $m_{\sigma}$ and the $f_{\pi}$, $f_K$ as input determine  the tree level QM model quartic couplings  $\lambda_1$, $\lambda_2$,~mass parameter  $m^2$,~$h_x$, $h_y$ and the coefficient $c$ of the t'Hooft determinant term for the $U_A(1)$ axial anomaly \cite{Rischke:00,Schaefer:09,grahl}.~The grand minima search  $\frac{\partial \Omega_{\rm RQM}}{\partial \x}= \frac{\partial \Omega_{\rm RQM}}{\partial \y}=0$ for the  Eq.~(\ref{grandRQM}) gives the  $T$ and  $\mu$ dependence of $ \x$ and $ \y$.
\begin{table*}[!htbp]
\caption{Chiral limit path is made by reducing the $\pi, \ K$ meson starred masses as  $\frac{m_{\pi}^*}{m_{\pi}}= \frac{m_{K}^*}{m_{K}}=\gamma \le 1$ such that $\frac{m_{\pi}^*}{m_{K}^*}=\frac{m_{\pi}}{m_{K}}$ where $m_{\pi}=138$ and $m_{K}=496$ MeV at physical point \cite{Schaefer:09}.~
The Eq.~(\ref{fpi}) and Eq.~(\ref{fk}) with the $m_{\pi}^*$, $m_{K}^*$  for any $\gamma$,~give the corresponding  $f_{\pi}, \ f_{K}$.~The experimental values $(m_{\eta},m_{\eta^{\prime}}) = (547.5,957.8)$ MeV give
the $M_{\eta}$:Expt=1103.22 MeV  while the ChpT expression in the  Eq.~(\ref{Meta}) gives $M_{\eta}$:ChPT=1035.55 MeV for the physical $m_{\pi}, m_{K}$   \cite{etamass}.~RQM$\lbrace \text{QM} \rbrace$ model nonstrange [strange] direction explicit chiral symmetry breaking strenths $ h^*_{x0} \lbrace h^*_{x} \rbrace $ [$ h^*_{y0} \lbrace h^*_{y}\rbrace$]  in  $\text{MeV}^3$ are respectively $(119.79)^3 \lbrace (120.98)^3 \rbrace $ [$(324.39)^3 \lbrace (336.65)^3 \rbrace $], $(75.78)^3 \lbrace (76.16)^3 \rbrace $ [$(197.66)^3 \lbrace (204.63)^3 \rbrace $] and $(62.4)^3 \lbrace (62.49)^3 \rbrace $ [$(162.07)^3 \lbrace (166.16)^3 \rbrace $]  for $\gamma=$ 1,0.25 and 0.14.~The other  RQM$\lbrace \text{QM} \rbrace$ model parameters are $\lambda_{20},c_{0},\lambda_{10} \ \text{and} \ m_{0}^2$ $\lbrace \lambda_{2},c,\lambda_{1} \ \text{and} \ m^2 \rbrace$.$\delta_{1}=7.283\times10^{-6};\delta_{2}=16.574\times10^{-6}$.} 
    \label{tab:table1}
\begin{tabular}{p{0.09\textwidth} p{0.08\textwidth}  p{0.08\textwidth} p{0.13\textwidth} p{0.14\textwidth} p{0.16\textwidth} p{0.14\textwidth} p{0.17\textwidth}} 
\hline
$ \gamma  $ & $f_{\pi} (\text{MeV})$ & $f_{K} (\text{MeV})$ & $M_{\eta}: (\text{MeV}$) & $\lambda_{20} \ \lbrace  \lambda_{2} \rbrace$ & $ c_{0} \ \lbrace  c \rbrace (\text{MeV})$& $\lambda_{10} \ \lbrace  \lambda_{1} \rbrace$ & $ m_{0}^{2} \ \lbrace  m^2 \rbrace  (\text{MeV}^2)$  \\
\hline
1 &92.9737 & 113.2635 & Expt=1103.22 & 31.19 $\lbrace 47.92 \rbrace$ & 7962.25 $\lbrace 4785.76 \rbrace$ & 3.56 $\lbrace -6.48 \rbrace$ & $(465.62)^2 \lbrace (495.28)^2 \rbrace $ \\
1 &92.9737&113.2635 &ChpT=1035.55 & 35.47  $\lbrace 54.46 \rbrace$ &7390.57  $\lbrace 3913.39 \rbrace$ &1.27 $\lbrace -9.87 \rbrace$ &$(447.71)^2 \lbrace (467.34)^2\rbrace $  \\
1 &92.9737&113.2635 & 650.5 & 90.81  $\lbrace 83.76 \rbrace$ &0  $\lbrace 0 \rbrace$ &-27.68 $\lbrace -25.06 \rbrace$ &$-(154.43)^2 \lbrace (311.53)^2\rbrace $  \\
0.5&92.7862&102.1041&865.99 &-12.95 $\lbrace 16.59 \rbrace$  &7705.88 $\lbrace 4241.24  \rbrace$ &20.19 $\lbrace 6.04 \rbrace$  &$(217.28)^2 \lbrace (283.01)^2\rbrace $ \\ 
0.3742 &91.510&97.723 &845.54 &-25.01 $\lbrace 5.86 \rbrace$  &7833.91 $\lbrace  4505.99 \rbrace$ &25.35 $\lbrace 11.05 \rbrace$ &$(157.17)^2 \lbrace (245.25)^2 \rbrace$ \\
$2.02\times10^{-4}$ &$88+\delta_{1}$ & $88+\delta_{2}$& 825.04 & -79.67 $\lbrace-47.2\rbrace$  & 8012.18 $\lbrace 5156.80 \rbrace$  & 46.21 $ \lbrace 32.39 \rbrace$  &$-(84.12)^2 \lbrace (182.89)^2 \rbrace $\\
\hline
\end{tabular}
\end{table*}
\begin{figure*}[htb]
\subfigure[\ Nonstrange condensate.]{
\label{fig1a} 
\begin{minipage}[c]{0.23\textwidth}
\centering \includegraphics[width=\linewidth]{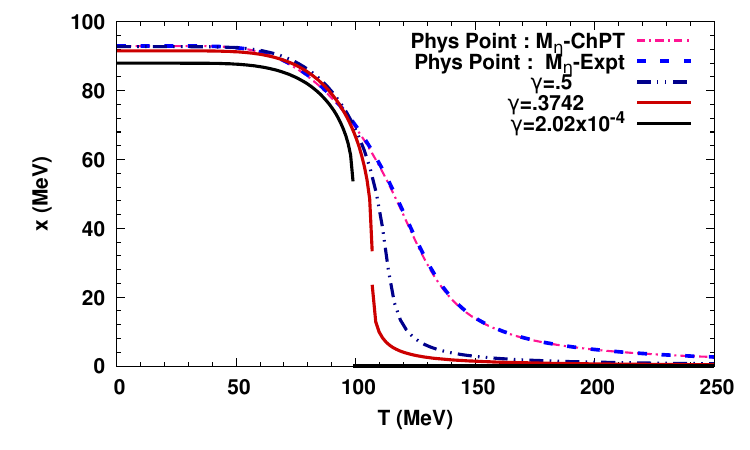}
\end{minipage}}
\hfill
\subfigure[\ Strange condensate.]{
\label{fig1b} 
\begin{minipage}[c]{0.23\textwidth}
\centering \includegraphics[width=\linewidth]{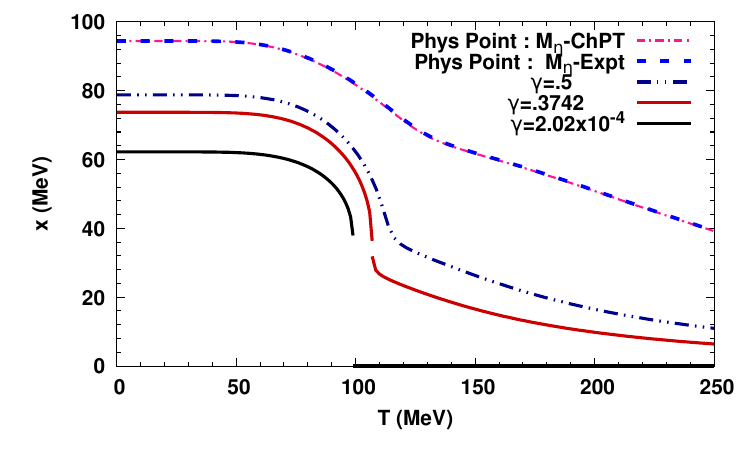}
\end{minipage}}
\hfill
\subfigure[\ Columbia plot: With $U_{A}$(1)]{
\label{fig1c} 
\begin{minipage}[c]{0.23\textwidth}
\centering \includegraphics[width=\linewidth]{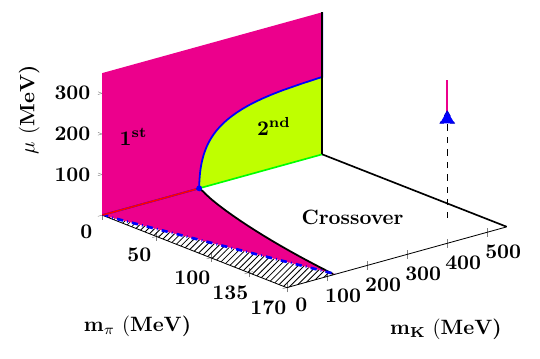}
\end{minipage}}
\hfill
\subfigure[\ Columbia plot: No $U_{A}$(1)]{
\label{fig1d} 
\begin{minipage}[c]{0.23\textwidth}
\centering \includegraphics[width=\linewidth]{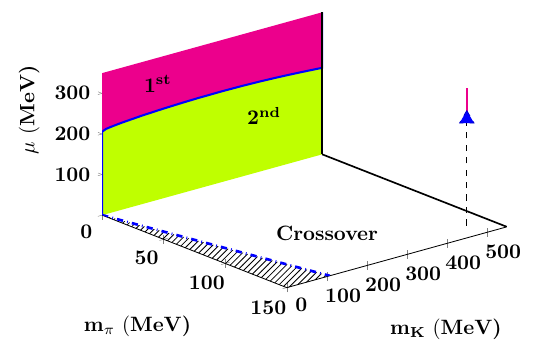}
\end{minipage}}
\vspace{-.3cm}
\caption{Temperature variations of $x$ and $y$ [(a) and (b)] at $\mu=0$ for the  ratio $\frac{m_{\pi}^*}{m_{\pi}}= \frac{m_{K}^*}{m_{K}}=\gamma=1,0.5,0.3742 \ \text{and} \ 2.02\times 10^{-4}.$~The $\mu-m_{K}$ and $m_{\pi}-m_{K}$ plane depict the respective  chiral transition for the $m_{\pi}=0$ and  $\mu=0$ [(c) with $U_{A}$(1) anomaly: $c_{0} \ne 0$, (d) without: $c_{0}=0$].~Solid black second order  critical line in [c] separates crossover from the first order region and solid blue line of tricritical points (starts  at the blue dot $m_{K}^{TCP}=242.6$ MeV in [c]),~separates the second  and first order regions in [c and d].~The $h_{y0}$ is negative in the dashed area near the $m_{\pi}$ axis.~The dash dot blue line depicts the strange chiral limit $h_{y0}=0$.~The vertical line at the physical point in the (c) and (d),~shows the crossover transition (black dashed line) ending at the critical end point (blue triangle) and first order transition (solid red line).~The value of $\sigma$ mass $m_{\sigma}=400$ MeV.   }
\vspace{-.5 cm}
\label{fig:mini:fig1} 
\end{figure*}

{\bf {Consistent path to chiral limit and ChPT scaling of $\bf f_{\pi},f_{K} \ \text{and} \ M_{\eta}^2=m_{\eta}^2+m_{\eta^{\prime}}^2$  }} 
Using the $ \mathcal{O}(\frac{1}{f^2})$ \cite{gasser, herrNPB, herrPLB, borasoyI, borasoyII} accurate results of the ChPT,
~the dependences of $f_{\pi},f_{K} $ and $M_{\eta}^2$ on the $m_\pi, m_K$ has the following form \cite{Herpay:05}.
\vspace {-.15 cm} 
\bqa 
\label{fpi}
\resizebox{0.9\hsize}{!}{$f_{\pi}=f-\frac{1}{f} [2\mu_{\pi}+\mu_{K}-4m_{\pi}^2 (L_{4}+L_{5})-8m_{K}^2 L_{4}]\;. $} \\  
\label{fk}
\resizebox{0.9\hsize}{!}{$f_{K}=f-\frac{1}{f}[ \frac{3}{4}(\mu_{\pi}+\mu_{\eta}+2\mu_{K})-4m_{\pi}^2 L_{4}  
 -4m_{K}^2(2L_{4}+L_{5}) ]\;$}.  \\ \nonumber
\label{Meta}
\nonumber
\resizebox{0.9\hsize}{!}{$M_{\eta}^2=2m_{K}^2-3v_{0}^{(2)}+2(2m_{K}^2+m_{\pi}^2)\
\{ 3v_{2}^{(2)}-v_{3}^{(1)} \}+$}  \\ \nonumber 
\resizebox{0.9\hsize}{!}{$ \frac{1}{f^2} \biggl[8v_{0}^{(2)}(2m_{K}^2+m_{\pi}^2)(3L_{4}+L_{5})+m_{\pi}^2(\mu_{\eta}-3\mu_{\pi})$}  \\ \nonumber
\resizebox{0.9\hsize}{!}{$-4m_{K}^2 \mu_{\eta}+\frac{16}{3}\biggl\{(6L_{8}-3L_{5}+8L_{7})(m_{\pi}^2-m_{K}^2)^2+$}\\
\resizebox{0.9\hsize}{!}{$2L_{6}(m_{\pi}^4-2m_{K}^4+m_{\pi}^2m_{K}^2)+L_{7}(m_{\pi}^2+2m_{K}^2)^2\biggr\} \biggr]\;.$}     \\ \nonumber
\eqa
The $\mu_{\text{\tiny{PS}}}=\frac{m_{\text{\tiny{PS}}}^2}{32 \pi^2} \ln(\frac{m_{\text{\tiny{PS}}}^2}{M_{0}^2})$ are chiral logarithms at scale $M_{0}$ and  $ m_{\text{\tiny{PS}}}$ is the leading order mass of the corresponding pseudoscalar octet meson. With the input $f_\pi$=93 MeV, $f_K$=113 MeV,  $m_\pi$=138 MeV, $m_K$=495.6 MeV,$m_\eta$=547.8 MeV and $M_{0}=4 \pi f_{\pi}\equiv 1168$ MeV, f=88 MeV,one obtains the chiral constants in the physical point  as $L_4=-0.7033 \times 10^{-3}$ and $L_5=0.3708 \times 10^{-3}$.~The determination of the constants is explained in Ref. \cite{Herpay:05}.~The other constants needed to find $ M_{\eta}^2 $ are $L_6=-0.3915 \times 10^{-3}$ , $L_7=-0.2272 \times 10^{-3}$, $L_8=0.511 \times 10^{-3}$, $v_{0}^{(2)}=-29.3 f^2$, $v_{3}^{(1)}=0.095$ and $v_{2}^{(2)}=-.1382$. 

{\bf  Path to chiral Limit and Columbia Plot} : The equations of motion $\frac{\partial \Omega_{vac}^{\rm RQM}}{\partial \x}=0= \frac{\partial \Omega_{vac}^{\rm RQM}}{\partial \y}$ using Eq.~(\ref{vacRQM})  give the renormalized explicit symmetry breaking strengths as $h_{x0}=m_{\pi,c}^2 \ f_{\pi} $ and $h_{y0}=(\sqrt{2}f_Km^2_{K,c}-\frac{f_{\pi}}{\sqrt{2}}m^2_{\pi,c})$ \cite{curvmass}.~For light chiral limit $h_{x0}=0$ while $h_{y0}=0$ defines the line of strange chiral limit.~The temperature variations of the light(strange) condensate $x(y)$ calculated from the two parameter sets $M_{\eta}:\text{Expt}$ and $M_{\eta}:\text{ChPT}$ are almost overlapping in   Fig.~\ref{fig1a}(Fig.~\ref{fig1b}).~The physical point ($\gamma$=1) crossover transition becomes sharper as the $\gamma$ is decreased to .5 and it turns second order (of the 3-d Ising universality class $Z_{2}$) at the critical $\gamma_{c}=.3795$ (not shown in the Fig) where $(m_{\pi}^*,m_{K}^*)=(52.38,~188.25)$ MeV.~The transition is first order for  $\gamma=.3742$ as $x(y)$ show a small gap in the Figs.~The $x(y)$ temperature dependences in Fig.~\ref{fig1a} (Fig.~\ref{fig1b}) show a very strong first order transition in the chiral limit  where a very small difference of $f_{\pi}=88+7.283 \times 10^{-6}$ from $f_{K}=88+16.574 \times 10^{-6}$ MeV ( for $\gamma=2.02\times10^{-4}$~with $ \lbrace m_{\pi}^{*},m_{K}^{*} \rbrace $=$\lbrace 0.1,.028 \rbrace $ MeV) enables  the calculation of parameteres and $h_{x0},h_{y0}$ are put to zero.~For the exact chiral limit the $f_{\pi}=f_{K}$,~the  $m_{\eta}=0$ and the heavy $m_{\eta^{\prime}} \equiv 825$ MeV signifies the $U_A(1)$ anomaly.~The problem of the loss of the SCSB when the $m_{\pi},m_{K} 
\rightarrow  0$  in the chiral limit for the smaller $m_{\sigma}=400-600$ MeV,~gets cured after using the ChPT scaling \cite{Herpay:05} of the  $m_{\pi},m_{K},M_{\eta}^2$.~This stands in contrast,~to the s-MFA QM model studies in Refs. \cite{Rischke:00,Schaefer:09} where the SCSB occurs for the unphysically large $\sigma$ mass $m_{\sigma} \ge 800$ MeV and also to the case of heuiristic parameter fixing in the fixed-$f_{\pi}$ scheme of the  nonperturbative FRG study of the Ref. \cite{Resch} where the SCSB is retained for the $m_{\sigma}=400-600$ MeV at each step of computation by adjusting the UV scale $\Lambda$ of the action heuiristically at every lower mass point in the Columbia plot.

The light chiral limit ($m_{\pi}=0$) chiral transition in the Fig.~(\ref{fig1c}) at $\mu=0$ ~is of second order for the  $m_{K} \ge 496$ MeV as in \cite{Resch} and also predicted in Ref.\cite{rob} while it is first order in the s-MFA QM model independent of the $m_{K}$ and  the $U_A(1)$ anomaly strength \cite{Schaefer:09}.~For $m_{K}<496$ MeV,~the  second order transition line of the O(4) universality,~ends at the blue dot of the tricritical point $m_{K}^{TCP}=242.6$ MeV in the Fig.~(\ref{fig1c}) where the transition turns first order and becomes stronger till the chiral limit is reached.~The chiral critical black line that separates the crossover from the first order region in the $m_{\pi}-m_{K}$ plane ($\mu=0$) for the non zero $m_{\pi}$,~terminates on the strange chiral limit line (the $h_{y0}=0$ axis is identical with $h_{y}=0$ in the blue dash dotted line) at the terminal $\pi$ mass $m_{\pi}^{t}\equiv168.39$ MeV  beyond which the transition is a smooth crossover everywhere.~For the  SU(3) symmetric chiral limit path in the Columbia plot when $m_{\pi}=m_{K}$,~the boundary of first order region ($\mu=0$) ends at the critical pion mass of $m_{\pi}^{c}=134.16$ MeV.~The s-MFA study of Ref. \cite{Schaefer:09} for the $m_{\sigma}=800$ MeV finds $m_{\pi}^{c}\equiv150$ MeV.~The chiral matrix model study using mean filed gives $m_{\pi}^{c}\equiv110$ MeV \cite{pisarski24}.

The first order region in our work with the $U_A(1)$ anomaly in Fig.\ref{fig1c},~is considerably larger than what is seen in the Fig.(3a) of the QM model e-MFA study done after switching off the quantum and thermal fluctuations of the mesons in the advanced FRG framework  of Ref.\cite{Resch} where one finds the $(m_{K}^{TCP},~m_{\pi}^{t},~m_{\pi}^{c}) \equiv (169,~110,~86)$ MeV for the $m_{\sigma}=530$ MeV while we find $(m_{K}^{TCP},~m_{\pi}^{t},~m_{\pi}^{c}) \equiv (242.6,~168.39,~134.16)$ MeV,~for the $m_{\sigma}=400$ MeV.~These critical values are about 6 to 8 MeV smaller if we take $m_{\sigma}=530$ MeV in our work.~Since the exact  on-shell renormalization of the parameters in RQM model,~gives a significantly stronger $U_{A}(1)$ anomaly strength $c$ and 
weaker light (strange) symmetry breaking strenght $h_{x}(h_{y})$,~the softening effect of the quark one-loop vacuum correction on the chiral transition is moderate.~Note that the strength $c$ and $h_{x}(h_{y})$ do not change either in the e-MFA FRG study or in the renormalized QM model potential in Ref.\cite{schafwag12} which gives similar effects as argued in
Ref.\cite{Resch}.~Also the  anomaly strength $c$ changes and becomes stronger (see Table I) for smaller masses in the Columbia plot due to the use of consistent ChPT scaling of $m_{\pi},m_{K},M_{\eta}$ for determing the QM/RQM model parameters.~Thus the softening effect of the quark one-loop vacuum fluctuation~on the chiral transition in the e-MFA FRG work using LPA,~looks overestimated  simliar to what has been shown is in our recent RQM model studies \cite{vkkr22, skrvkt24} for the e-MFA QM model works in Ref.\cite{schafwag12, chatmoh1, vkkr12, vkkt13} which use the minimal subtraction scheme to dimensionally regularize the vacuum divergences and curvature meson masses to fix the model parameters.~The full FRG study finds very low $m_{\pi}^{c}\equiv17$ MeV for which Pisarski et. al. \cite{pisarski24} note that this is partly because the approximation used in Ref. \cite{Resch} is known to overestimate the mesonic fluctuations that tend to soften the transition \cite{PawlRen}.~Note that  the Ref.\cite{fejoHastuda} has also cautioned that LPA in FRG studies,~completely neglects the wave function renormalization which might change the final result.

The size of the robust first order region near the chiral limit increases on increasing $\mu$.~Since the chiral critical surface has a positive curvature \cite{forcrd2},~its intersection with the dashed black vertical line of crossover transition at the physical mass point,~marks the existence of a critical end point in a solid blue arrow in the $\mu-T$ plane of Fig.~\ref{fig1c}.~The positive slope of the chiral tricritcal line (which lies above the second order and below the first order region) starting at $m_{K}^{TCP}=242.6, \ \mu=0$ MeV,~keeps on decreasing for larger $\mu$ and $ m_{K}$ and it shows near saturation as  quite a small slope of .049 is obtained when the $\mu=188.25$ MeV at $m_{K}=500$ MeV becomes $\mu=190.7$ MeV at $m_{K}=550$ MeV.~This tricritcal line is expected to be connected to  the tricritical point of the two flavor chiral limit \cite{hjss} at some higher value of $\mu$ and $m_{K}$.

The $m_{\pi}-m_{K}$ plane of Fig.~(\ref{fig1d}) has no first order region without the $U_A(1)$ anomaly unlike the findings in \cite{rob,Schaefer:09}.~This is the effect of quark one-loop vacuum fluctuation similar to the e-MFA:FRG study in \cite{Resch}.
~The second order  chiral transition for every $m_{K}$ when $m_{\pi}=0$ and $\mu<\mu_{c}$,~turns first order in the $\mu-m_{K}$ plane at the origin $\mu_{c}=204.6$ MeV, $m_{K}=0$ (chiral limit) of a tricritcal line  whose small positive slope upto $\mu_{c}=223.2,m_{K}=240$ MeV,~becomes constant for $m_{K}=240-350$ MeV and then decreases very slowly till the $\mu_{c}=213.2,~m_{K}=550$ MeV.~The solid blue arrow marks the citical end point $\mu_{CEP}=262.8$ MeV at the physical mass.



\begin{thebibliography}{99}
\bibitem{curvmass}
The light and strange symmetry breaking strenghts in the QM model are $h_{x}=m_{\pi}^2 \ f_{\pi} $ and $h_{y}=(\sqrt{2}f_Km^2_{K}-\frac{f_{\pi}}{\sqrt{2}}m^2_{\pi})$.
The pion and kaon curvature masses $m_{\pi,c}$ and $m_{K,c}$ as derived in Ref. \cite{vkkr22} are :
$m_{\pi,c}^2=m_{\pi}^2\lbrace 1-\frac{N_{c}g^2}{4\pi^2} m_{\pi}^2 \mathcal{C}^{\prime}(m^2_\pi,m_u)  \rbrace$ \ \ ; 
$m_{K,c}^2=m_{K}^2\lbrace 1-\frac{N_{c}g^2}{4\pi^2} (a-b+c)\rbrace$ where $a=\mathcal{C}(m^2_\pi,m_u)+m^2_\pi\mathcal{C}^{\prime}(m^2_\pi,m_u)$,
$b=(1-\frac{(m_s-m_u)^2}{m_K^2})\mathcal{C}(m^2_K,m_u,m_s) $ and 
$c=(1-\frac{f_\pi}{f_K})\frac{m^2_u-m_s^2+2m_s^2\ln(\frac{m_s}{m_u})}{m^2_K} $.
\bibitem{etamass}
~With the input of $M_{\eta}=\sqrt{(m_{\eta}^2+m_{\eta^{\prime}}^2)}=1103.22$ MeV ( termed as $M_{\eta}:\text{Expt}$) for the experimental  $(m_{\eta},m_{\eta^{\prime}}) = (547.5,957.8)$ MeV,~the parameters of the QM model give $(m_{\eta},m_{\eta^{\prime}}) = (538.54,962.85)$ MeV in the ouput while the renormalized parameters of the RQM model give $(m_{\eta},m_{\eta^{\prime}}) = (527.82,968.76)$ MeV in the ouput.~The results are consistent because the calculated $M_{\eta}$ in the output  of the QM (RQM) model is the same as in the  input.~The ChPT expression  in Eq.(\ref{Meta}) gives a slightly smaller value of $M_{\eta}=1035.55$  ( termed as $M_{\eta}:\text{ChPT}$ ) MeV  at the physical point and with this $M_{\eta}$ in the input,~the QM model parameters give $(m_{\eta},m_{\eta^{\prime}}) = (528.76,890.79)$ MeV
in the ouput while the renormalized parameters of the RQM model give $(m_{\eta},m_{\eta^{\prime}}) = (520.62,895.17)$ MeV in the ouput, here also the  output $M_{\eta}$ is same as in the input.
\end{thebibliography}

\bibliographystyle{apsrmp4-2}

\end{document}